# Zero modes of the Dirac operator in three dimensions

C. Adam*

School of Mathematics, Trinity College, Dublin 2

B. Muratori**, C. Nash***

Department of Mathematical Physics, National University of Ireland, Maynooth

## Abstract

We investigate zero modes of the Dirac operator coupled to an Abelian gauge field in three dimensions. We find that the existence of a certain class of zero modes is related to a specific topological property precisely when the requirement of finite Chern–Simons action is imposed.

*)email address: adam@maths.tcd.ie, adam@pap.univie.ac.at

**)email address: bmurator@fermi1.thphys.may.ie

***)email address: cnash@stokes2.thphys.may.ie

# 1   Introduction

Zero modes for Fermions for the Dirac operator $\slashed{D}_A = \slashed{\partial} - i\slashed{A}$ are of importance in many places in quantum field theory. They are the ingredients for the computation of the index of the Dirac operator and play a key rôle in understanding anomalies. In Abelian gauge theories, which is what we are concerned with here, they affect crucially the behaviour of the Fermion determinant $\det(\slashed{D}_A)$ in quantum electrodynamics. The nature of the QED functional integral depends strongly on the degeneracy of the bound zero modes. In three dimensions little is known about such Fermion bound states, the first examples being only obtained in 1986 in the paper of [1]. In our paper we want to further investigate this problem of zero modes of the Abelian Dirac operator in three dimensional Euclidean space (i.e., the Pauli operator).

It should be emphasized here that the problem of zero modes of the Pauli operator, in addition to being interesting in its own right, has some deep physical implications. The authors of [1] were mainly interested in these zero modes because in an accompanying paper [2] it was proven that one-electron atoms with sufficiently high nuclear charge in an external magnetic field are unstable if such zero modes of the Pauli operator exist.

Further, there is an intimate connection between the existence and number of zero modes of the Pauli operator for strong magnetic fields on the one hand, and the nonperturbative behaviour of the three dimensional Fermionic determinant (for massive Fermions) in strong external magnetic fields on the other hand. The behaviour of this determinant, in turn, is related to the paramagnetism of charged Fermions, see [3, 4]. So, a thorough understanding of the zero modes of the Pauli operator is of utmost importance for the understanding of some deep physical problems, as well.

In addition, it is speculated in [4] that the existence and degeneracy of zero modes for $QED_3$ may have a topological origin as it does in $QED_2$—cf. [4] for details and an account of the situation for $QED_{2,3,4}$. The results of the present paper tend to support such a point of view.

Our paper is organized as follows.
In Section 2, we review the simple example of a zero mode that was given in [1], and we provide some more examples of zero modes of a similar type. In Section 3, we show, for a whole class of zero modes, that their existence is related to a specific, associated "topological number" being an odd integer. This result will follow from the conditions of square integrability of the zero mode and finiteness of the Chern–Simons action. If only the weaker condition of square integrable zero modes and magnetic fields is imposed, we find a broader class of possible zero modes. In Section 4, a generalization to higher angular momentum is discussed.

# 2   Some examples

First we want to briefly review a specific solution that was given in [1]. Let $\Psi$ be a two-component, square integrable spinor on $\mathbf{R}^3$ and $A_i$ a gauge field leading to a square integrable field strength $F_{ij} = A_{j,i} - A_{i,j}$ (further $x = (x_1, x_2, x_3)$, $r = |x|$, and $\sigma_i$ are the Pauli matrices). The authors of [1] observed that a solution of the Dirac equation

$$-i\slashed{\partial}\Psi(x) \equiv -i\sigma_i\partial_i\Psi(x) = A_i(x)\sigma_i\Psi(x) \tag{1}$$

could be easily obtained from a solution of the simpler equation

$$-i\slashed{\partial}\Psi = h\Psi \tag{2}$$

for some scalar function $h(x)$. In this case the corresponding gauge field that obeys the Dirac equation (1) together with the spinor (2) is given by

$$A_i = h\frac{\Psi^\dagger \sigma_i \Psi}{\Psi^\dagger \Psi}. \tag{3}$$

In addition, they gave the following explicit example

$$\Psi = (1+r^2)^{-\frac{3}{2}}(\mathbf{1} + i\vec{x}\vec{\sigma})\Phi_0 \tag{4}$$

where $\Phi_0$ is the constant unit spinor $\Phi_0 = (1,0)^{\mathrm{T}}$. The spinor (4) obeys

$$-i\slashed{\partial}\Psi = \frac{3}{1+r^2}\Psi \tag{5}$$

and is, therefore, a zero mode for the gauge field

$$\vec{A} = \frac{3}{1+r^2}\frac{\Psi^\dagger \vec{\sigma}\Psi}{\Psi^\dagger \Psi} = \frac{3}{(1+r^2)^2}\begin{pmatrix} 2x_1x_3 - 2x_2 \\ 2x_2x_3 + 2x_1 \\ 1 - x_1^2 - x_2^2 + x_3^2 \end{pmatrix}. \tag{6}$$

(The geometrical behaviour of the gauge field (6) is shown in Figs. 1, 2. The integral curves of this gauge field are closed circles that lie on tori, and they wrap once around each direction of the torus (Fig. 1a). Any two different curves are linked exactly once (see Fig. 1b for two curves on different tori).)

Next we want to give some more examples of solutions of the type (2), (3), that may be obtained by simple generalizations of the solution (4), (6). From these examples we will find that a zero mode exists when a certain "topological number" is an odd integer, and we will show in the next section that this feature holds true for a whole class of zero modes.

Some first examples of solutions may be constructed from the following ansatz

$$\Psi^{(l)} = (1+r^2)^{-(\frac{3}{2}+l)}[(1 + \sum_{n=1}^{l} a_n r^{2n})\mathbf{1} + \sum_{n=0}^{l} b_n r^{2n}\mathbf{X}]\Phi_0 \tag{7}$$

where

$$\mathbf{X} \equiv ix_j\sigma_j. \tag{8}$$

The sums are restricted by square integrability of $\Psi^{(l)}$, and the choice $a_0 = 1$ just fixes an arbitrary normalization. With the help of the relations

$$\mathbf{X}^2 = -r^2 \mathbf{1}, \quad x_j \partial_j \mathbf{X} = \mathbf{X}, \quad i\sigma_j \partial_j \mathbf{X} = -3 \cdot \mathbf{1} \tag{9}$$

one easily finds, e.g., for $l = 1$

$$-i\partial\!\!\!/\Psi^{(1)} = -i\partial\!\!\!/(1+r^2)^{-\frac{5}{2}}[(1+a_1 r^2)\mathbf{1} + (b_0 + b_1 r^2)\mathbf{X}]\Phi_0 \tag{10}$$

$$= \frac{3b_0}{(1+r^2)^{\frac{7}{2}}}[(1 + \frac{5b_1 - 2b_0}{3b_0}r^2)\mathbf{1} + (\frac{5 - 2a_1}{3b_0} + \frac{a_1}{b_0}r^2)\mathbf{X}]\Phi_0 \tag{11}$$

and the proportionality condition (2) can be met by comparing coefficients of powers of $r^2$. The resulting system of equations is linear in all coefficients except $b_0$, for which it is of fourth order. Actually, it is of second order in $b_0^2$, because the sign flip $b_i \to -b_i$ just corresponds to a parity transformation $x_i \to -x_i$. One of the two solutions for $b_0^2$ just reproduces the simplest solution $l = 0$, (4), by producing a common factor $1 + r^2$ in the numerator of (10), the other solution is new,

$$\Psi^{(1)} = (1+r^2)^{-\frac{5}{2}}[(1 - \frac{5}{3}r^2)\mathbf{1} + (\frac{5}{3} - r^2)\mathbf{X}]\Phi_0 \tag{12}$$

$$-i\partial\!\!\!/\Psi^{(1)} = \frac{5}{1+r^2}\Psi^{(1)}. \tag{13}$$

This pattern repeats itself for higher $l$. The condition (2) always leads to a system of equations for the $a_i$, $b_i$ with $l + 1$ different solutions (up to parity). $l$ of these solutions just reproduce the lower $\Psi^{(k)}$, $k < l$, whereas one solution produces a new zero mode, $\Psi^{(l)}$. Explicitly, the coefficients $a_i$, $b_i$ for $\Psi^{(l)}$ read

$$a_i = (-1)^l b_{|i-l|}$$

$$\Big(\prod_{j=0}^{i}(2j+3)\Big)(i!)b_i = (-1)^i \Big(\prod_{k=0}^{i-1}(l-k)\Big)\Big(\prod_{h=0}^{i}(2l+3-2h)\Big) \tag{14}$$

and, with the abbreviations $A = \sum_{n=0}^{l} a_n r^{2n}$, $(a_0 \equiv 1)$, and $B = \sum_{n=0}^{l} b_n r^{2n}$ the corresponding gauge field reads

$$\vec{A}^{(l)} = \frac{2l+3}{(1+r^2)(A^2 + B^2 r^2)}\begin{pmatrix} 2B^2 x_1 x_3 - 2AB x_2 \\ 2B^2 x_2 x_3 + 2AB x_1 \\ A^2 + B^2(x_3^2 - x_1^2 - x_2^2) \end{pmatrix} \tag{15}$$

(for the behaviour of $\vec{A}^{(1)}$ see Fig. 3). All these zero modes behave like $\Psi^\dagger \Psi \sim r^{-4}$ for $r \to \infty$, as well as the corresponding gauge fields, $\vec{A}^2 \sim h^2 \sim r^{-4}$.

Some more examples of zero modes of the type (2) may be found by the following observation. The simplest zero mode (4) may be expressed like

$$\Psi = g(r)U\Phi_0, \quad g(r) = \frac{1}{1+r^2}, \quad U = \frac{1}{(1+r^2)^{\frac{1}{2}}}(\mathbf{1} + \mathbf{X}) \tag{16}$$

where $g$ is a scalar function (depending only on $r$) and $U$ is an $SU(2)$ matrix. So one may wonder whether there are zero modes for higher powers of $U$,

$$\Psi^{(n)} = gU^n\Phi_0 \tag{17}$$

where $g$ is a scalar that has to be determined. E.g., for $n = 2$ one finds

$$\Psi^{(2)} = gU^2\Phi_0 = g(1+r^2)^{-1}[(1-r^2)\mathbf{1} + 2\mathbf{X}]\Phi_0 \tag{18}$$

and, with a little bit of algebra ($' \equiv \partial/\partial r^2$)

$$-i\not{\partial}g(r^2)(1+r^2)^{-1}[(1-r^2)\mathbf{1} + 2\mathbf{X}]\Phi_0 =$$

$$\frac{2}{(1+r^2)^2}[(2r^2(1+r^2)g' + (3+r^2)g)\mathbf{1} + (2g - (1+r^2)(1-r^2)g')\mathbf{X}]\Phi_0$$

$$\stackrel{!}{=} hg(1+r^2)^{-1}[(1-r^2)\mathbf{1} + 2\mathbf{X}]\Phi_0. \tag{19}$$

Comparison of the coefficients of $\mathbf{1}$ and $\mathbf{X}$ leads to two first order linear differential equations for the function g, which have to be proportional. This determines $h$ and, in turn, g (we abbreviate $u \equiv r^2$)

$$1) \quad g' = \frac{(1-u)(1+u)h - 2(3+u)}{4u(1+u)}g$$

$$2) \quad g' = \frac{2 - (1+u)h}{(1-u)(1+u)}g$$

$$\Rightarrow \quad h = \frac{2(3-u)}{(1+u)^2}$$

$$g' = \frac{-4}{(1+u)^2}g, \quad g = \exp(\frac{4}{1+u})$$

and we find

$$\Psi^{(2)} = \frac{e^{\frac{4}{1+r^2}}}{1+r^2}[(1-r^2)\mathbf{1} + 2\mathbf{X}]\Phi_0 \tag{20}$$

$$-i\not{\partial}\Psi^{(2)} = \frac{2(3-r^2)}{(1+r^2)^2}\Psi^{(2)}. \tag{21}$$

This formal solution $\Psi^{(2)}$ is not square integrable and, therefore, is not an acceptable zero mode. However, the same ansatz for the third power of $U$,

$$\Psi^{(3)} = gU^3\Phi_0 = \frac{g}{(1+r^2)^{\frac{3}{2}}}[(1-3r^2)\mathbf{1} + (3-r^2)\mathbf{X}]\Phi_0 \tag{22}$$

leads to a square integrable zero mode,

$$\Psi^{(3)} = \frac{e^{\frac{8}{(1+r^2)^2} - \frac{8}{1+r^2}}}{(1+r^2)^{\frac{5}{2}}}[(1-3r^2)\mathbf{1} + (3-r^2)\mathbf{X}]\Phi_0 \tag{23}$$

$$-i\not\partial\Psi^{(3)} = \frac{9r^4 - 14r^2 + 9}{(1+r^2)^3}\Psi^{(3)}. \tag{24}$$

Zero modes for higher powers $U^n$ ($n = 4, 5, \ldots$) of $U$ may be computed in an analogous fashion, and one again finds that for odd powers of $U$ there exist square integrable zero modes, whereas for even powers the formal solutions are not square integrable. This may lead to the conjecture that this is a general property, i.e., odd powers of $U$ in (17) always lead to square integrable zero modes, whereas even powers do not. In the next scetion we will show that this is indeed the case.

Further, one might speculate that this different behaviour is related to different geometrical or topological properties of even and odd powers of $U$. So, let us briefly discuss the geometry of the matrices $U^n$. The first power

$$U = (1+r^2)^{-\frac{1}{2}}(\mathbf{1} + ir\vec{n}\vec{\sigma}) \tag{25}$$

$$n_j = \frac{x_j}{r}, \quad \vec{n}^2 = 1 \tag{26}$$

has the following properties

$$U(r=0) = \mathbf{1}, \quad U(r=\infty) = i\vec{n}\vec{\sigma}, \tag{27}$$

i.e., it behaves similarly to an $SU(2)$ monopole and tends to the "hedgehog" shape $i\vec{n}\vec{\sigma}$ for $r \to \infty$ (actually this hedgehog corresponds to the identity map $S^2(r=\infty) \to S^2$(unit vector $\vec{n}$) with winding number 1). The matrix $U^2$

$$U^2 = (1+r^2)^{-1}[(1-r^2)\mathbf{1} + 2ir\vec{n}\vec{\sigma}] \tag{28}$$

has the properties

$$U^2(r=0) = \mathbf{1}, \quad U^2(r=1) = i\vec{n}\vec{\sigma}, \quad U^2(r=\infty) = -\mathbf{1} \tag{29}$$

i.e., it is defined on $\mathbf{R}^3$ compactified to $S^3$ (actually, it corresponds to a map $S^3 \to S^3$ with winding number 1), and, therefore, it is similar to a Skyrmion. In addition, it is equal to the hedgehog at the sphere of radius 1.

For higher powers of $U$ this pattern repeats itself. As $\lim_{r\to\infty} U^n = (i\vec{n}\vec{\sigma})^n$, odd powers of $U$ tend to the hedgehog $\pm i\vec{n}\vec{\sigma}$ for $r \to \infty$, i.e., they are of the monopole type. Even powers tend to $\pm \mathbf{1}$, i.e., they are of the Skyrmion type.

## 3  A general class of solutions

Here we want to study a class of $SU(2)$ matrices, and we want to show that, depending on the imposed integrability conditions (see below), these $SU(2)$ matrices do provide zero modes precisely when they are of the monopole type (i.e., they tend to the hedgehog configuration for $r \to \infty$). For this purpose we use a parametrization of $SU(2)$ matrices that was used in [5] for a discussion of Skyrmions. They use the ansatz for a class of $SU(2)$ matrices $U$ (in polar coordinates $r, \theta, \phi$)

$$U = \exp(if(r)\vec{n}(\theta,\phi)\vec{\sigma}). \tag{30}$$

Here the profile function $f$ depends on $r$, and the unit vector $\vec{n}$ depends on $\theta, \phi$. Therefore, $\vec{n}$ defines a map $S^2 \to S^2$ with integer winding number.

Via a stereographic projection, the coordinate two-sphere $(\theta, \phi)$ may be mapped onto the complex plane $\mathbf{C}$ with coordinate $z$. Explicitly, the map is $z = \tan(\theta/2)e^{i\phi}$. In this new coordinate, a class of unit vectors $\vec{n}(\theta, \phi)$ may be expressed by rational maps $z \to R(z) = \frac{p(z)}{q(z)}$, where $p, q$ are co-prime polynomials, and the degree of the map (the degree of $p$ or $q$, whichever is higher) equals the winding number of the map $(\theta, \phi) \to \vec{n}(\theta, \phi)$. Explicitly, $\vec{n}(R(z))$ is

$$\vec{n}(R(z)) = \frac{1}{1+|R|^2}\begin{pmatrix} 2\mathrm{Re}R(z) \\ 2\mathrm{Im}R(z) \\ 1-|R(z)|^2 \end{pmatrix}. \tag{31}$$

In the sequel we will restrict to the simplest rational map $R(z) = z$ (the identity map $S^2 \to S^2$)

$$\vec{n}(z) = \frac{1}{1+|z|^2}\begin{pmatrix} 2\mathrm{Re}z \\ 2\mathrm{Im}z \\ 1-|z|^2 \end{pmatrix} = \begin{pmatrix} \sin\theta\cos\phi \\ \sin\theta\sin\phi \\ \cos\theta \end{pmatrix} = \frac{1}{r}\begin{pmatrix} x_1 \\ x_2 \\ x_3 \end{pmatrix}. \tag{32}$$

In the ansatz (30) for $U$ we further assume, without restriction, that $f(0) = 0$, i.e., $U(r=0) = \mathbf{1}$. Then $f(\infty) = 2k\frac{\pi}{2}$, $k \in \mathbf{Z}$ corresponds to a Skyrmion-type $SU(2)$ field with baryon number (i.e., $S^3 \to S^3$ winding number) $k$. Further, $f(\infty) = (2k+1)\frac{\pi}{2}$ corresponds to a monopole-type $SU(2)$ field.

Therefore, we now want to prove that the ansatz

$$\Psi = g(r)\exp(if(r)\vec{n}(z)\vec{\sigma})\Phi_0 = g(r)[\cos f(r)\mathbf{1} + i\sin f(r)\vec{n}(z)\vec{\sigma}]\Phi_0$$

$$=: [C(r)\mathbf{1} + iS(r)\vec{n}(z)\vec{\sigma}]\Phi_0 \tag{33}$$

$$C(r) := g(r)\cos f(r), \quad S(r) := g(r)\sin f(r) \tag{34}$$

may provide square integrable zero modes of the type $-i\not{\partial}\Psi = h(r)\Psi$ for monopole-type $U$ but not for Skyrmion-type $U$.

*Remark:* Before continuing, we want to point out that all our examples of Section 2 belong to this ansatz (33) (with $\vec{n}$ given by (32)). Indeed, all examples (e.g., (7), (17)) may be written like $\Psi = (a(r)\mathbf{1} + b(r)\mathbf{X})\Phi_0$, where $a(r), b(r)$ are some rational functions (for (17) this is true because $\mathbf{X}^2 = -r^2\mathbf{1}$). Using $\mathbf{X} = ir\vec{n}\vec{\sigma}$, this is equal to ansatz (33) with $a(r) = C(r)$, $rb(r) = S(r)$. Further, higher powers $U^k$ of an SU(2) matrix $U$ (as we used in (17)) are of the type (30) if $U$ is, because $U^k = \exp(ikf(r)\vec{n}(z)\vec{\sigma})$. I.e., $U^k$ is computed by substituting $kf(r)$ instead of $f(r)$ for the profile function in (30).

To simplify the computation, we use the fact that the matrix $U$ acts on the spinor $\Phi_0 = (1,0)^{\mathrm{T}}$, i.e., not all components of $U$ actually occur in $\Psi$. This is best achieved by introducing the matrices

$$P_0 = \frac{1}{2}(\mathbf{1} + \sigma_3) = \begin{pmatrix} 1 & 0 \\ 0 & 0 \end{pmatrix}, \quad P_0 \begin{pmatrix} 1 \\ 0 \end{pmatrix} = \begin{pmatrix} 1 \\ 0 \end{pmatrix}$$

$$P_1 = \frac{1}{2}(\sigma_1 - i\sigma_2) = \begin{pmatrix} 0 & 0 \\ 1 & 0 \end{pmatrix}, \quad P_1 \begin{pmatrix} 1 \\ 0 \end{pmatrix} = \begin{pmatrix} 0 \\ 1 \end{pmatrix}$$

$$\sigma_1 P_0 = P_1, \quad \sigma_2 P_0 = iP_1, \quad \sigma_3 P_0 = P_0$$

$$\sigma_1 P_1 = P_0, \quad \sigma_2 P_1 = -iP_0, \quad \sigma_3 P_1 = -P_1 \tag{35}$$

The spinor (33) may be rewritten like

$$\Psi = [(C(r) + iS(r)n_3(z))P_0 + iS(r)n_+(z)P_1]\Phi_0 \tag{36}$$

where

$$n_+ = n_1 + in_2, \quad x_+ = x_1 + ix_2, \quad \partial_+ = \frac{1}{2}(\partial_1 - i\partial_2). \tag{37}$$

Acting with $-i\slashed{\partial}$ on $\Psi$ gives

$$-i\slashed{\partial}\Psi = [\partial_3(-iC + Sn_3) + 2\partial_+(Sn_+)]P_0\Phi_0 + [2\partial_-(-iC + Sn_3) - \partial_3(Sn_+)]P_1\Phi_0. \tag{38}$$

Now we have to introduce polar coordinates $(r, z, \bar{z})$,

$$x_+ = \frac{2rz}{1 + z\bar{z}}, \quad \partial_+ = \frac{\bar{z}}{1 + z\bar{z}}\partial_r + \frac{1}{2r}(\partial_z - \bar{z}^2\partial_{\bar{z}})$$

$$x_3 = \frac{r(1 - z\bar{z})}{1 + z\bar{z}}, \quad \partial_3 = \frac{1 - z\bar{z}}{1 + z\bar{z}}\partial_r - \frac{1}{r}(z\partial_z + \bar{z}\partial_{\bar{z}}) \tag{39}$$

and use (32) for our simple choice of $\vec{n}(z)$,

$$n_+ = \frac{2z}{1 + z\bar{z}}, \quad n_3 = \frac{1 - z\bar{z}}{1 + z\bar{z}} \tag{40}$$

to find, after some algebra (here the prime denotes derivative w.r.t. $r$, *not* $r^2$),

$$-i\slashed{\partial}\Psi = [-iC'n_3 + S' + \frac{2}{r}S]P_0\Phi_0 - in_+C'P_1\Phi_0$$

$$\stackrel{!}{=} h(r)[(C + iSn_3)P_0 + iSn_+P_1]\Phi_0. \tag{41}$$

This leads to the two differential equations

$$S' + \frac{2}{r}S = hC \tag{42}$$

$$-C' = hS \tag{43}$$

which we rewrite in terms of the functions

$$t(r) := \tan f(r) = \frac{S(r)}{C(r)} \tag{44}$$

$$g(r) = \sqrt{C^2(r) + S^2(r)}. \tag{45}$$

We arrive at

$$t' = h(1 + t^2) - \frac{2}{r}t \quad \Rightarrow \quad h = (1 + t^2)^{-1}(t' + \frac{2}{r}t) \tag{46}$$

$$g' = -\frac{2}{r}\frac{t^2}{1 + t^2}g. \tag{47}$$

Now we assume that a function $t$ is given with the properties

$$\lim_{r \to 0} t(r) \sim r^{\alpha_0}, \quad \lim_{r \to \infty} t(r) \sim r^{\alpha_\infty}. \tag{48}$$

Regularity of $h$ and the corresponding magnetic field $\vec{B} = \vec{\partial} \times \vec{A}$ at $r = 0$ requires

$$\lim_{r \to 0} t(r) \sim cr + O(r^{2+\epsilon}), \quad \epsilon \geq 0 \quad \Rightarrow \quad \alpha_0 = 1 \text{ or } \alpha_0 \geq 2. \tag{49}$$

Concerning the behaviour for $r \to \infty$, we want to discuss two conditions separately. As a stronger condition, we require that the Chern–Simons action shall be finite (i.e., the Chern–Simons density integrable), in addition to square integrability of the magnetic field and of the zero mode. Explicitly, the Chern–Simons density for our ansatz (33) is, after some computation (we use the notation of differential forms here, i.e., $A = A_i dx^i$, etc.)

$$AdA = \frac{2h^2}{1 + t^2}(\frac{t}{r}(1 + n_3^2) + t'n_+n_-)r^2 dr \frac{4}{(1 + a^2)^2}da\, d\phi \tag{50}$$

where

$$a := z\bar{z}, \quad \sin\theta d\theta = \frac{4}{(1 + a^2)^2}da \tag{51}$$

and

$$\int AdA = 4\pi(2\pi - 4)\int_0^\infty dr \frac{rh^2 t}{1 + t^2} + 4\pi(-2\pi + 8)\int_0^\infty dr \frac{r^2 h^2 t'}{1 + t^2} \tag{52}$$

where $h$ is given by (46). Integrability of $AdA$ requires $\alpha_\infty \neq 0$, as may be checked easily. In addition, square integrability of $g$ (i.e., of the zero mode) requires $\alpha_\infty \geq 0$ and, therefore, we end up with the condition

$$\alpha_\infty > 0. \tag{53}$$

Next, we have to relate this asymptotic behaviour of $t$ to the properties of the matrix $U$, (30). For illustrative purposes, let us first do it for the explicit example (22) (the third power $U^3$ of the specific simplest matrix $U$, (16)). There we find

$$t(r) = \tan f(r) = \frac{r(3 - r^2)}{1 - 3r^2} \tag{54}$$

and therefore

$$\tan f(0) = 0 \quad \ldots \quad f(0) = 0, \quad \tan f(\frac{1}{\sqrt{3}}) = \infty \quad \ldots \quad f(\frac{1}{\sqrt{3}}) = \frac{\pi}{2}$$

$$\tan f(\sqrt{3}) = 0 \quad \ldots \quad f(\sqrt{3}) = \pi, \quad \tan f(\infty) = \infty \quad \ldots \quad f(\infty) = \frac{3\pi}{2} \qquad (55)$$

showing that the "topological number"

$$N := (f(\infty) - f(0))\frac{2}{\pi} = 3 \qquad (56)$$

is an odd integer in this case, and, therefore, the corresponding $SU(2)$ matrix is of the monopole type. Now this consideration may be immediately generalized to general $t$ with the behaviour (49), (53). From (49) it follows that

$$t(0) = \tan f(0) = 0 \quad \ldots \quad f(0) = 0. \qquad (57)$$

From (53) we conclude

$$t(\infty) = \tan f(\infty) = \infty \quad \ldots \quad f(\infty) = (2k+1)\frac{\pi}{2}, \ k \in \mathbf{Z} \qquad (58)$$

showing that square integrable zero modes have odd "topological number" $N = 2k+1$ and are, therefore, of the monopole type, if the additional condition of finite Chern–Simons action is imposed. On the other hand, when $\alpha_\infty < 0$ (leading to a non-square integrable zero mode), then

$$t(\infty) = \tan f(\infty) = 0 \quad \ldots \quad f(\infty) = 2k\frac{\pi}{2}, \ k \in \mathbf{Z}. \qquad (59)$$

Therefore, Skyrmion-type $SU(2)$ matrices with even "topological number" lead to non-square integrable, formal solutions of the Pauli equation.

This is what we wanted to prove.

Now we want to impose the weaker condition that the magnetic field $\vec{B}$ be square integrable. This does not give a condition on $\alpha_\infty$ at all. Further, square integrability of $g$ just requires $\alpha_\infty \geq 0$, where we have already discussed the case $\alpha_\infty > 0$. So let us investigate the case

$$\alpha_\infty = 0, \quad t(\infty) = t_\infty = \text{const.} \qquad (60)$$

a little bit closer. Square integrability of $g$ requires

$$2\frac{t_\infty^2}{1+t_\infty^2} > \frac{3}{2} \quad \Rightarrow \quad t_\infty^2 > 3$$

$$\Rightarrow \quad f_\infty \in (\frac{\pi}{3} + k\pi, \frac{2\pi}{3} + k\pi), \quad k \in \mathbf{Z}. \qquad (61)$$

Therefore, now a broader class of zero modes is allowed, where the $SU(2)$ matrix $U$, (30), may tend to a mixture of monopole and Skyrmion for $r \to \infty$,

$$U(r = \infty) \sim \cos f_\infty \mathbf{1} + \sin f_\infty i\vec{n}\vec{\sigma}, \quad \sin^2 f_\infty > 3\cos^2 f_\infty. \tag{62}$$

In this case, a quantity that generalizes the "topological number" $N$, (56), to non-integer values, may be computed from the matrix $U$, (30),

$$N(f_\infty) := \frac{1}{12\pi^2} \int \mathrm{tr}(U^\dagger dU)^3 = \frac{4}{\pi} \int_0^\infty \sin^2 f(r) f'(r) dr$$

$$= \frac{4}{\pi} \int_0^{f_\infty} \sin^2 f\, df = \frac{2}{\pi}(f_\infty - \frac{1}{2}\sin 2f_\infty) \tag{63}$$

($f(0) = 0$). Obviously, it reduces to the integer $\frac{2}{\pi} f_\infty$ for pure monopoles and Skyrmions.

## 4 Higher angular momentum

The authors of [1] observed that, in addition to their simplest solution (4), they could find similar solutions to equ. (2) with higher angular momentum. Using instead of the constant spinor $\Phi_0 = (1,0)^\mathrm{T}$ the spinor

$$\Phi_{l,m} = \begin{pmatrix} \sqrt{l+m+1/2}\, Y_{l,m-1/2} \\ -\sqrt{l-m+1/2}\, Y_{l,m+1/2} \end{pmatrix} \tag{64}$$

(where $m \in [-l-1/2,\, l+1/2]$ and $Y$ are spherical harmonics), they found the solutions

$$\Psi_{l,m} = r^l(1+r^2)^{-l-\frac{3}{2}}(\mathbf{1} + \mathbf{X})\Phi_{l,m} \tag{65}$$

$$\vec{A}_{l,m} = (2l+3)(1+r^2)^{-1}\frac{\Psi_{l,m}^\dagger \vec{\sigma} \Psi_{l,m}}{\Psi_{l,m}^\dagger \Psi_{l,m}}. \tag{66}$$

This may be immediately generalized to our ansatz (33). For the spinor

$$\Psi_{l,m} = (C(r)\mathbf{1} + iS(r)\vec{n}\vec{\sigma})\Phi_{l,m} \tag{67}$$

we find, by a computation that is similar to the one leading to (42), (43)

$$S' + \frac{2+l}{r}S = hC \tag{68}$$

$$C' - \frac{l}{r}C = -hS \tag{69}$$

or

$$h = (1+t^2)^{-1}(t' + \frac{2(1+l)}{r}t) \tag{70}$$

$$g' = (-\frac{2+l}{r}\frac{t^2}{1+t^2} + \frac{l}{r}\frac{1}{1+t^2})g. \tag{71}$$

Again, we discuss the condition of finite Chern–Simons action first. There, the conditions on $t$ remain the same, (49), (53). Further, $g$ behaves like

$$\lim_{r\to 0} g(r) \sim r^l, \quad \lim_{r\to\infty} g(r) \sim r^{-l-2} \tag{72}$$

where the first condition is just the usual angular momentum barrier.

If only square integrability of the magnetic field is required, then again $\alpha_\infty = 0$, $t(\infty) = t_\infty$ is possible, and we find as the condition for square integrability of $g$

$$t_\infty^2 > \frac{3+2l}{1+2l}. \tag{73}$$

## 5 Summary

For a whole class of zero modes (ansatz (33) with condition (47)) we have shown that their existence is related to a topological quantization condition (the topological number (63) is an odd integer) precisely when the additional requirement of finite Chern–Simons action of the corresponding gauge field (85) is imposed (geometrically, this topological quantization condition implies that the associated $SU(2)$ matrix (30) is of the monopole type). This result clearly points towards a topological origin of the whole problem, and it may also be of some interest to Chern–Simons QFT.

From the above construction it seems that we related the existence of a square integrable zero mode to a topological property of this zero mode (i.e., of the $SU(2)$ matrix $U$) rather than of the gauge field. This is not necessarily true, however. Remember that all our zero modes (33) are of the specific type

$$\Psi = gU\Phi_0, \quad -i\slashed{\partial}\Psi = h\Psi, \quad A_i = \frac{h}{2}\text{tr}\, U^\dagger \sigma_i U \sigma_3 \tag{74}$$

i.e., $A_i$ is related to the matrix $U$ in a simple algebraic manner. This is true even for the scalar function $h$, which is determined algebraically by $t$ (and $t'$) and does not depend on $g$, see (46). Therefore, the $SU(2)$ matrix $U$ of ansatz (30) uniquely determines the gauge field $A_i$ via (46), (85), and it may well be that it is ultimately the topology of $A_i$ that determines the existence of zero modes.

Still, there remain many open questions. E.g., our ansatz (33) (or (67)) only provides one zero mode per gauge field, and it would be interesting to find examples with more than one zero mode (if they exist). Further, one would, of course, like to understand, which topological property determines the number and existence of zero modes in the general case. These questions are subject to further investigation.

It should be mentioned here that there is an index theorem for odd-dimensional open spaces like $\mathbf{R}^3$, however, for a slightly different field contents. The Callias index theorem [6] was formulated for a Dirac operator $\slashed{D}_{A,\Phi}$ in a space of odd dimensions in a Yang–Mills–Higgs background $(A, \Phi)$, and there it is actually the nontrivial, monopole-like behaviour of the Higgs field $\Phi$ for $r \to \infty$ that accounts for a nonzero index (i.e., difference of the number of zero modes of $\displaystyle{\not}D_{A,\Phi}$ and $\displaystyle{\not}D_{A,\Phi}^\dagger$). Whether this index theorem may be of some relevance in the present context is not clear yet.

Finally, we want to remark that an apparently obvious generalization to (33) does not work. In (33) the simplest possible unit vector $\vec{n}(\theta,\phi)$, (32), is used (corresponding to the rational map $R(z) = z$). One may wonder whether new solutions may be obtained by allowing for other unit vectors in (33), e.g., the ones with winding number $m$ corresponding to the rational maps $R(z) = z^m$, $m \geq 2$. The answer is no. The point is that every zero mode has to obey the condition

$$\vec{\partial}\Psi^\dagger \vec{\sigma}\Psi = 0. \tag{75}$$

If one inserts ansatz (33) with a higher winding $\vec{n}$ into this condition, one immediately realizes that it cannot be fulfilled as long as both $f$ and $g$ only depend on $r$.

# 6  Acknowledgments

The authors gratefully acknowledge useful conversations with M. Fry. CA is supported by a Forbairt Basic Research Grant. BM gratefully acknowledges a fellowship from the Training and Mobility of Researchers scheme (TMR no. ERBFMBICT983476).

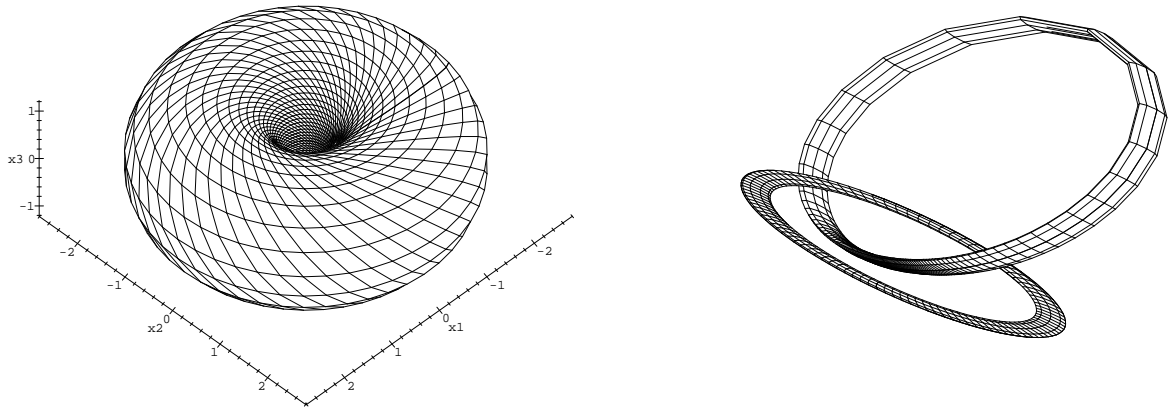

Figure 1: Integral curves of the vector field $\vec{A}^{(0)}$ linking exactly once for different parameter values

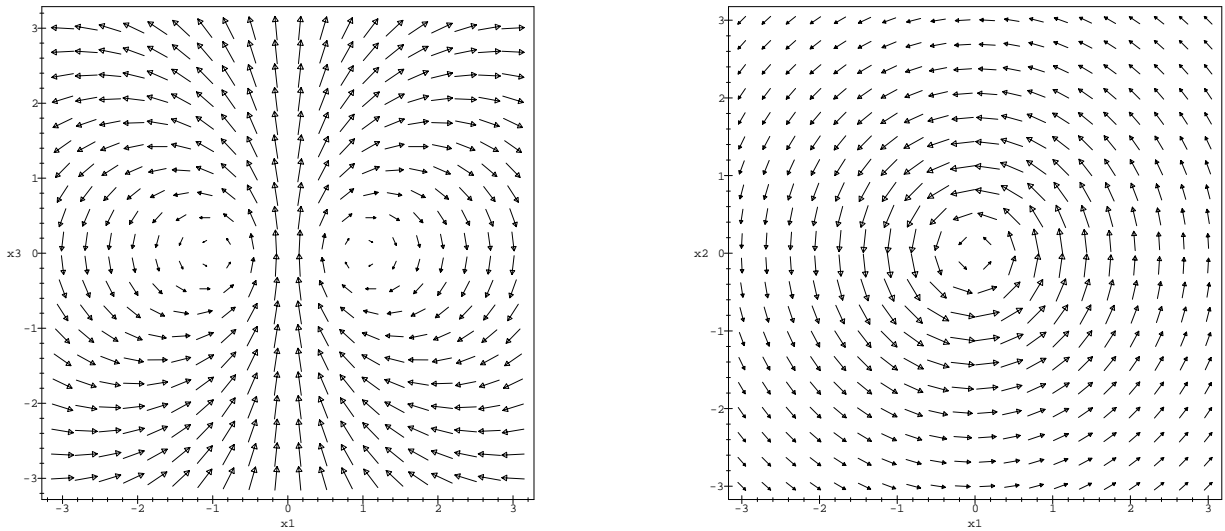

Figure 2: Cross sections $(x_2 = 0)$ and $(x_3 = 0)$ of the vector field $\vec{A}^{(0)}$

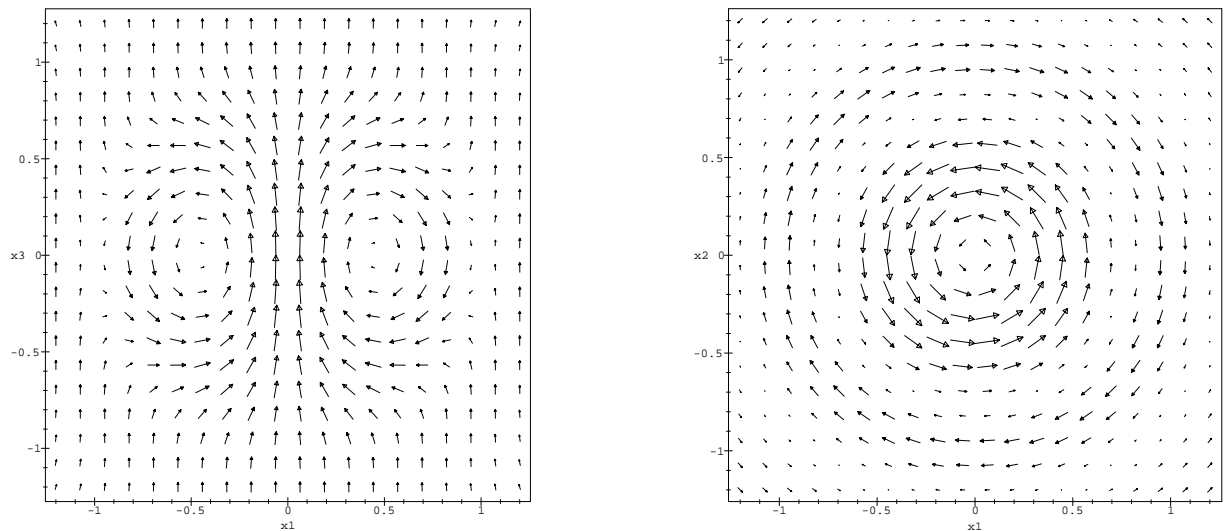

Figure 3: Cross sections $(x_2 = 0)$ and $(x_3 = 0)$ of the vector field $\vec{A}^{(1)}$